\documentclass[structabstract]{aa}

\usepackage{graphicx,epsfig,fancyhdr,rotating,amsmath,natbib,amssymb}

\newcommand{\degree}{\ensuremath{^\circ}}
\usepackage{txfonts}

\begin{document}

\title{Spectroscopic Observations of Propagating Disturbances in a Polar Coronal Hole:
       Evidence of Slow Magneto-acoustic Waves}

\author{G.~R.~Gupta	\inst{1}
	\and
	L.~Teriaca	\inst{1}
	\and
	E.~Marsch	\inst{1,2} 
	\and
        S.~K.~Solanki	\inst{1,3}
	\and
	D.~Banerjee	\inst{4}
	}

\institute{Max-Planck-Institut f\"{u}r Sonnensystemforschung (MPS), 37191  
	  Katlenburg-Lindau, Germany
	  \email{gupta@mps.mpg.de}
	  \and
	  Institute for Experimental and Applied Physics (IEAP), Christian Albrechts University at Kiel, 24118 Kiel, Germany
	  \and
	School of Space Research, Kyung Hee University, Yongin, Gyeonggi 446-701, Korea
    \and
	Indian Institute of Astrophysics, Bangalore 560034, India
    }


\abstract
{}
{We focus on detecting and studying quasi-periodic propagating features that have been interpreted both in 
terms of slow magneto-acoustic waves and of high speed upflows.}
{We analyze long duration spectroscopic observations of the on-disk part of the south polar 
coronal hole taken on 1997 February 25 by the SUMER spectrometer aboard SOHO. 
We calibrated the velocity with respect to the off-limb region and obtain time--distance maps in 
intensity, Doppler velocity and line width. We also perform a cross correlation analysis on different time series
curves at different latitudes. We study average spectral line profiles at the roots of propagating disturbances and
 along the propagating ridges, and perform a red-blue asymmetry analysis.}
{We find the clear presence of propagating disturbances in intensity and 
Doppler velocity with a projected propagation speed of about $60\pm 4.8$~km~s$^{-1}$ and 
a periodicity of $\approx$14.5~min. To our knowledge, this is the first simultaneous detection of
 propagating disturbances in intensity as well as in Doppler velocity in a coronal hole.
 During the propagation, an intensity enhancement is associated with a blue-shifted Doppler velocity. 
These disturbances are  clearly seen in intensity also at higher latitudes (i.e. closer to the limb),
 while disturbances in Doppler velocity becomes faint there. The spectral line profiles averaged along
 the propagating ridges are found to be symmetric, to be well fitted by a single Gaussian, and have
 no noticeable red-blue asymmetry.}
{Based on our analysis, we interpret these disturbances in terms of propagating slow magneto-acoustic waves.}


\titlerunning{Spectroscopic Observations of Propagating Disturbances} 
\authorrunning{Gupta et al.}

\maketitle

\section{Introduction}

Magnetohydrodynamic (MHD) waves are important candidates for the heating of the solar 
corona and the acceleration of the fast solar wind \citep{2009LRSP....6....3C,2010LRSP....7....4O}. 
Recently, \citet{2011Natur.475..477M} reported the ubiquitous presence of outward-propagating transverse 
oscillations with amplitudes of nearly 20~km~s$^{-1}$ and period between 100 and 500~s throughout the quiescent
 atmosphere which they claim to be energetic enough to accelerate the fast solar wind and heat the quiet corona. 
Numerous reports on the detection of various MHD wave modes in different parts 
of the solar atmosphere are present in the literature 
\citep{2005LRSP....2....3N,2007SoPh..246....3B,2011SSRv..158..267B}.
Depending on the modes of propagation, these waves can give a variety of 
signatures in various spectral line profile parameters 
\citep{2010ApJ...721..744K}. The propagation of slow magneto-acoustic 
waves in the upper solar atmosphere gives oscillations in intensity as well as 
in the Doppler velocity with propagation speed less than or equal to the 
sound speed \citep{1978JPhB...11..613B,1983SoPh...88..179E}.
Based on these observational signatures, intensity oscillations with periods 
between 3 and 30~min observed in polar coronal holes 
\citep[e.g.,][]{1998ApJ...501L.217D,1999ApJ...514..441O,
2009A&A...499L..29B,2011A&A...528L...4K} and in active region loops 
\citep[e.g.,][]{1999SoPh..186..207B,2000A&A...355L..23D,2002SoPh..209...61D,
2003A&A...404L...1K,2006A&A...448..763M,2008SoPh..252..321M,
2009ApJ...697.1674M,2011A&A...526A..58S} were interpreted in terms of slow 
magneto-acoustic waves. From spectroscopic studies, similar oscillations were 
detected in intensity and Doppler velocity 
\citep[e.g.,][]{2001A&A...368.1095O,
2003A&A...406.1105W,2003A&A...402L..17W,2009A&A...493..251G,
2009ApJ...696.1448W,2009A&A...503L..25W,2010ApJ...713..573M,2011ApJ...737L..43N}  
and were, again, mostly interpreted as slow magneto-acoustic waves.   

Recently, with imaging observations, propagating intensity disturbances with speed 
between 50 and 200~km~s$^{-1}$ detected in active regions 
\citep[e.g.,][]{2007Sci...318.1585S,2008ApJ...678L..67H,2009ApJ...706L..80M}
 and in coronal holes \citep[e.g.,][]{2010A&A...510L...2M,
2011ApJ...736..130T} were alternatively interpreted as recurrent high-speed upflows.
 At the same time, spectroscopic observations show
blue wing asymmetries in transition region and coronal emission lines 
\citep[e.g.,][]{2009ApJ...701L...1D,2009ApJ...707..524M,2010ApJ...722.1013D,
2011ApJ...727....7M,2011ApJ...727L..37T} which were attributed to the presence of 
ubiquitous high-speed upflows in the observed regions. These upflows could be
recurrent phenomena and would cause quasi-periodic oscillations in intensity, 
Doppler velocity and width \citep{2010ApJ...722.1013D}. 
Thus, interpretation of the observed oscillations in terms of recurrent 
high-speed upflows have challenged the wave interpretation. 

Coronal holes are the source region of the high-speed solar wind and are associated with
 rapidly expanding open magnetic fields \citep[see review by ][]{2009LRSP....6....3C}.
\citet{2005ApJS..156..265C} and \citet{2007ApJS..171..520C} describe the role of MHD waves in the
 heating of the corona and acceleration of the solar wind in coronal holes. Moreover, the presence of
 high speed outflows at the base of coronal holes can provide heated mass to the corona and the fast solar wind
\citep{2010A&A...510L...2M,2011ApJ...736..130T}. Thus, detection of propagating disturbances in coronal
hole would contribute to the understanding of the heating of the gas in the coronal hole and
 the acceleration of the fast solar wind, either in terms of propagating waves or high speed flows
 or a  superposition of both. In order to distinguish between these scenarios, good spectroscopic
 observations of these features are needed. 

In this paper, we focus on very long duration \textquoteleft sit--and--stare\textquoteright\ 
mode spectroscopic observations of a polar coronal hole in order to identify
 propagating disturbances within our field of view and study their nature. We also 
look for the source and physical events responsible for these disturbances and the 
observed periodicity. In addition, we tested the applicability of the Red-Blue ($R-B$) asymmetry 
analysis to our dataset.

\section{Data reduction and analysis}
\label{sec:obs}

\begin{figure*}[htb]
 \centering
\includegraphics[width=11cm]{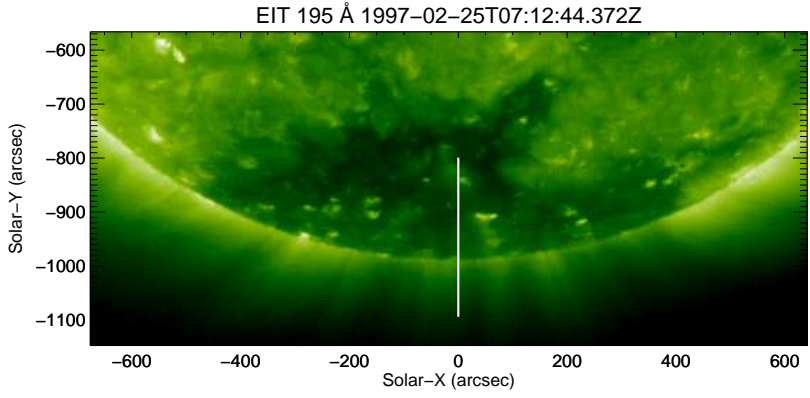}
\caption{EIT/SOHO 195~\AA\ context image of the southern polar coronal hole showing 
the location of the SUMER slit.}
\label{fig:context}
\end{figure*}

The data analyzed here were obtained on 1997 February 25 in the south polar coronal 
hole in \textquoteleft sit--and--stare\textquoteright\ mode with the Solar Ultraviolet 
Measurement of Emitted Radiation \citep[SUMER,][]{1995SoPh..162..189W} aboard SOHO. 
Two third of the SUMER slit was placed on-disk whereas the remaining part was off-limb.  
 Observations started at 00:00 UT and 
ended at around 14:00 UT with an exposure time of 60~s, recording the Ne~{\sc viii} 
770~\AA\ and N~{\sc iv} 765~\AA\ spectral lines in the first order of diffraction on detector B. 
Fig.~\ref{fig:context} shows 
the location of the SUMER slit in an image of the south polar coronal hole taken in
the 195~\AA\ passband with the Extreme ultraviolet Imaging Telescope 
\citep[EIT,][]{1995SoPh..162..291D} aboard SOHO. 
\citet{2009ApJ...704.1385S} also analyzed the same dataset to identify and 
characterize jet events. 

\begin{figure*}[hbt]
 \centering
\includegraphics[width=13cm]{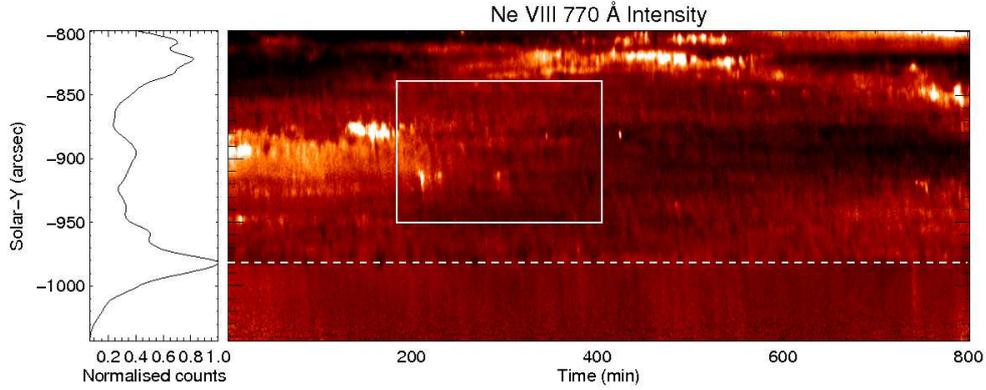}
\caption{The left panel shows the normalized time-average intensity
variation along the slit. The right panel depicts the intensity variation along solar-$Y$ with time, which is
normalized by the time-averaged intensity variation along the slit. The white dashed line marks the location
of maximum limb brightening, whereas the continuous box indicates the region chosen for detailed analysis.}
\label{fig:xt_org}
\end{figure*}

\begin{figure*}[htb]
 \centering
\includegraphics[width=15cm]{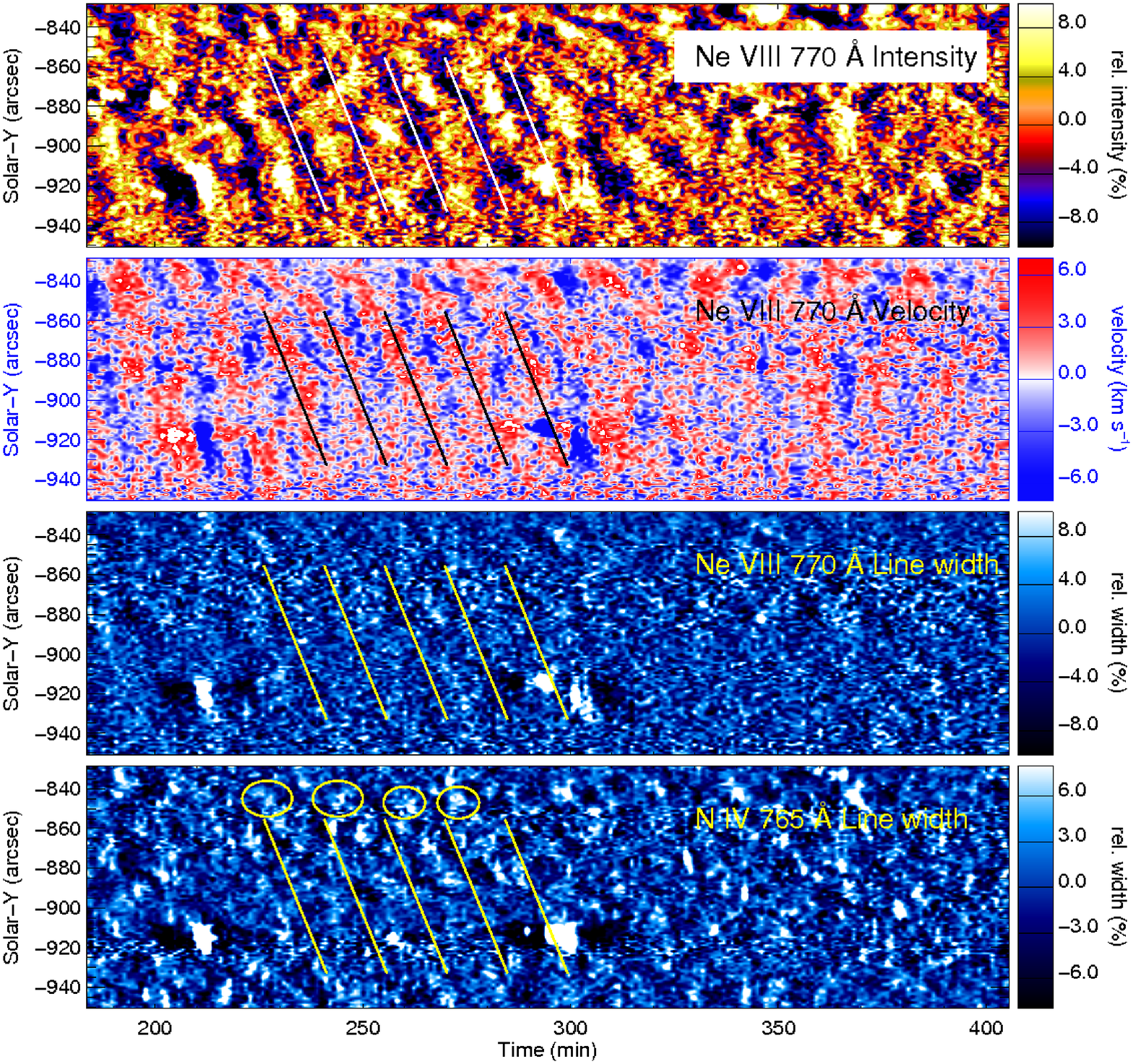}
\caption{Most interesting part of the enhanced distance--time map of the intensity (top panel),
 Doppler velocity (upper middle panel) and line width (lower middle panel) of the 
 Ne~{\sc viii} 770~\AA\ spectral line, and map of the line width (lower panel) of the 
 N~{\sc iv} 765~\AA\ spectral line. Propagating disturbances moving towards the limb
 are clearly visible in the top and middle 
panels and have a propagation speed of $60 \pm 4.8$~km~s$^{-1}$. 
White and black continuous lines in the top two panels are drawn to trace 
propagating disturbances with a period of 14.5~min. The bottom panel shows significant enhancements 
in line width at lower heights in the solar atmosphere, possibly indicating either the occurrence of 
unresolved small-scale reconnections or explosive events, or a local increase in turbulence
 at the base of the propagating disturbances (marked with ellipses).}  
\label{fig:xt_proc}
\end{figure*}

Data were first decompressed, corrected for dead-time, response inhomogeneities (flat-field),
 local-gain depression and for geometrical distortion (de-stretching), 
using the routines provided in SolarSoft. Single Gaussian fits were
applied to obtain the line peak, Doppler-shift and width of each spectrum, thus
obtaining distance-time ($Y-T$) maps of all three line parameters. 
These maps show horizontal stripes due to residuals in the flat-field correction.
Additionally, the Doppler-shift $Y-T$ map shows slow variations with time due to
changes in the instrument temperature during the 14~h duration of
 the observations and a trend along the slit due to residuals from the distortion correction.
Intensity and line width maps were divided by the normalized time-averaged profile along the slit. 
In the case of the Ne~{\sc viii} 770~\AA\ Doppler-shift map, the correction requires more care so as 
to retain the variations along the slit that are of solar origin (e.g.,
center-to-limb variations). The medium and large-scale
variations along the slit due to residuals in the distortion correction was obtained 
from far-off-limb spectra in cold lines (pure stray-light), which were taken two days before at 
nearly the same detector position and then subtracted from the data. 
Small-scale (few pixels) residual effects were removed by further dividing the 
velocity map by the normalized, high-pass-filtered, time-averaged 
profile along the slit. 
Finally, to convert the Doppler-shifts into absolute Doppler velocities,
we have set the average Doppler velocity in the off-limb part (between 
solar-$Y\approx-990\arcsec$ and $-1050\arcsec$) to zero in each frame \citep[LOS motions are expected to average
out above the limb in an optically thin plasma,][]{1976ApJ...205L.177D}. 
This procedure also removes the temporal variation 
along the wavelength scale due to the changes in the instrument temperature. 
The Doppler velocity calibrated in this way gives an average outflow speed of about 
2.5~km~s$^{-1}$ in Ne~{\sc viii} in the on-disk polar coronal hole, consistent with
previous studies  at similar latitudes
\citep[e.g.,][]{2000A&A...353..749W}. Fig.~\ref{fig:xt_org} shows the resultant $Y-T$ map
 obtained in normalized Ne~{\sc viii} 770~\AA\ intensity. 
The left panel displays the normalized intensity variation along the slit.
Many propagating disturbances (slanted ridges) in the on-disk part of the $Y-T$ map can be seen in the 
right panel of Fig.~\ref{fig:xt_org}. We identified one such region
where these propagating disturbances are very clearly seen in most of the line 
parameters. This region is outlined with a white box on the intensity $Y-T$ map of 
Fig.~\ref{fig:xt_org}. For the remaining part of this paper we will focus our attention 
on this region.
To see these propagating disturbances more clearly, one needs to remove any low-frequency
changes in intensity, Doppler velocity and width from the time series. 
To do this, we have
smoothed all the three $Y-T$ maps using the boxcar size of [3,20] in pixel units and then subtracted it 
from the original maps. To get the relative change in intensity and width, we divided 
these quantities by the smoothed map, obtaining the resultant enhanced  $Y-T$ maps 
 as shown in Fig.~\ref{fig:xt_proc}. Thus,
the applied processing can be summarized as, \textit{$\delta$L(t,y) = (L(t,y) -- \={L}(t,y))/ \={L}(t,y)}  and
 \textit{$\delta$v(t,y) = v(t,y) -- \={v}(t,y)} where \textit{L} stands for either intensity or line width,
 \textit{v} for velocity, and \textit{\={L}} and \textit{\={v}} are the smoothed parameters.
 
The solar rotation at $X\approx 0\arcsec$, at date of observation, goes from $2.8\arcsec$/hour at $Y\approx -840\arcsec$
 to $1\arcsec$/hour at $Y\approx -940\arcsec$. These values are comparable with, or smaller than,
 the SUMER spatial resolution of $2\arcsec$\ \citep{1997SoPh..170..105L} ensuring that the same location
 is observed for an adequate period of time.

\section{Results and discussion}
\label{sec:result}

Distance--time maps of intensity and velocity (Fig.~\ref{fig:xt_proc}) clearly reveal the presence of
 propagating disturbances 
 with 5\% to 10\% variations in intensity and 3 to 6~km~s$^{-1}$ changes in the Doppler velocity.
To our knowledge, this is the first simultaneous detection of propagating disturbances in intensity
 and Doppler velocity in a coronal hole region, in contrast to earlier detections, which were
restricted to active regions \citep{2009A&A...503L..25W,2011ApJ...727L..37T,2011ApJ...737L..43N}. 
The slanted bright and dark ridges (blue and red, in case of Doppler velocity) in the top two
 panels of Fig.~\ref{fig:xt_proc} are found to move towards the limb (i.e. outward from the Sun) with a
projected propagation speed of $60 \pm 4.8$~km~s$^{-1}$. In both these panels, white and black continuous 
lines indicate the expected trajectories of disturbances propagating at $\approx60$~km~s$^{-1}$ and
 with a 14.5~min periodicity. Clearly, enhancements in intensity are associated with blue-shifts 
(i.e. motion away from the Sun) whereas reductions in intensity are associated with red-shifts (inward motion).
The speeds measured here are comparable to previously reported ones from Ne~{\sc viii} 770~\AA\
 in a polar coronal hole, but off-limb \citep{2009A&A...499L..29B}. 
Comparing all the $Y-T$ maps, we deduce that these propagating disturbances are best visible 
in the zero-th moment of the spectral line (line intensity) followed closely by their visibility 
 in Doppler velocity, but are less clearly evident in line width. In the velocity $Y-T$ map,
these propagating disturbances are seen better at lower latitudes and become fainter at 
 higher latitudes (closer to the limb). This result explains why previous studies at higher latitudes
 and off-limb in polar coronal holes with the same Ne~{\sc viii}
770~\AA\ spectral line did not detect such disturbances in Doppler velocities, but only in intensity
\citep{2009A&A...499L..29B,2010ApJ...718...11G}.   

\begin{figure*}[htb]
 \centering
\includegraphics[width=15cm]{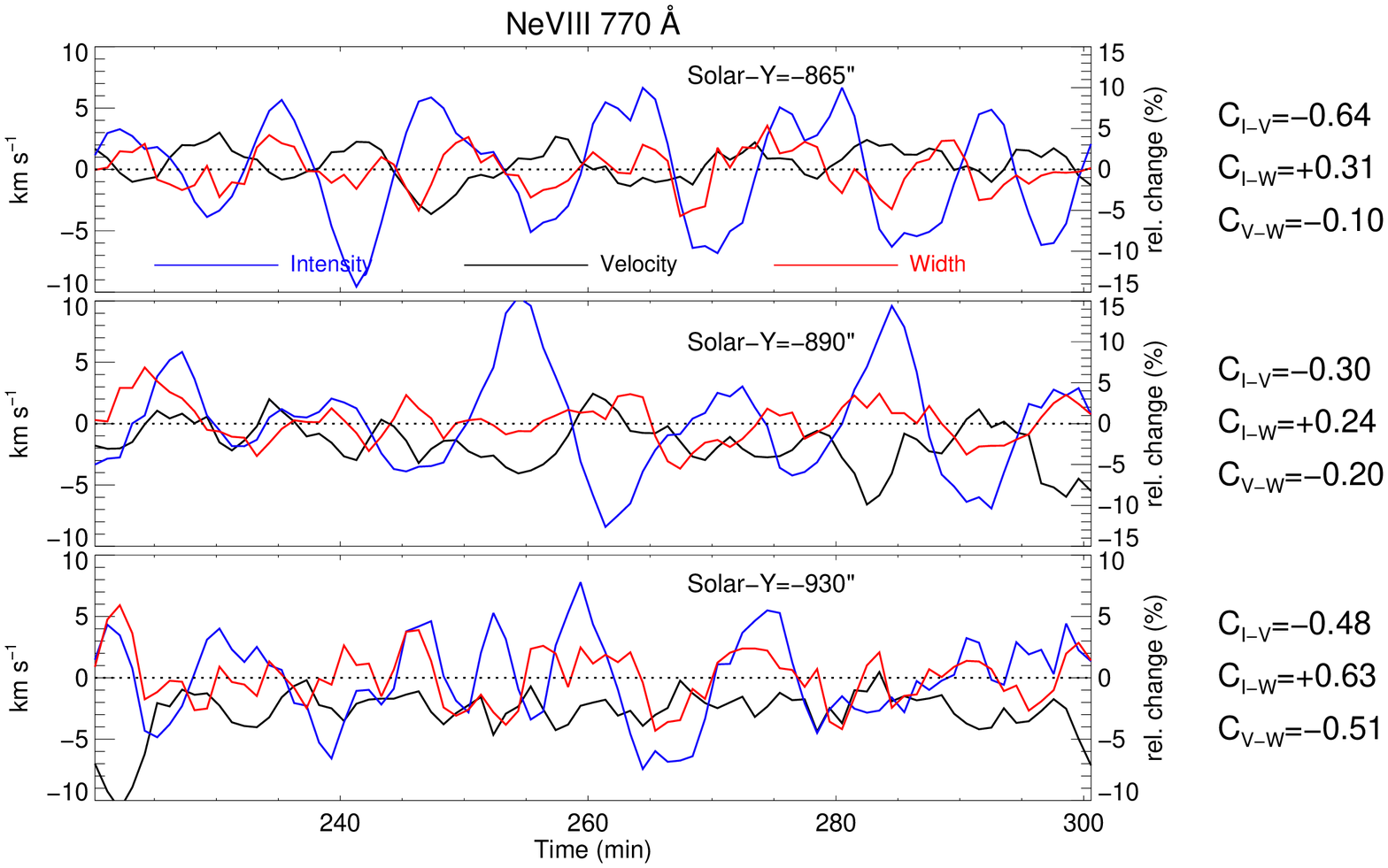}
\caption{Time series of Ne~{\sc viii} 770~\AA\ spectral line intensity, Doppler velocity and 
 width, which are obtained from $Y-T$ maps at three different locations averaged over 3 spatial 
 pixels. The measured Doppler-shifts are shown in km~s$^{-1}$, while the other parameters are in
 relative percentage change. The zero-velocity line is also plotted to discriminate between red
 and blue-shifts. Linear correlation coefficients between different non-detrended time series curves
 are provided for each latitudes on the right side of each frame. The 95\% and 99\% confidence level for 
linear correlation coefficients are about 0.217 and 0.284, respectively.}
\label{fig:lcurve}
\end{figure*}


To demonstrate the typical phase relationships between different spectral line parameters,
the time series, as obtained for all the three parameters,
 are plotted in Fig.~\ref{fig:lcurve} at three different latitudes.
The anti-correlation between intensity and Doppler velocity and the correlation between intensity and 
Doppler width are best seen in the top panel, but partly persist also at higher latitudes (lower panels). 
The strongest anti-correlation is between intensity and velocity (above 99\% confidence level at almost
all latitudes), followed by the positive correlation between intensity and line width (above 95\% 
confidence level at almost all latitudes). However, here we wish to point out that error bars associated
 with line width from Gaussian fitting are larger than the amplitude of their variations, hence,
 no major conclusion can be drawn from the line width part. Also, there is no correlation observed between
 Doppler width and Red-Blue asymmetry maps obtained from the Ne~{\sc viii} 770~\AA\ spectral profiles.

Recently, \citet{2009ApJ...696.1448W} and \citet{2010ApJ...724L.194V} showed that 
 the phase at which  an outward propagating slow magneto-acoustic wave produces  
a blue-shift of the line profile, coincides with a density (i.e. intensity) enhancement,
 in agreement with Figs.~\ref{fig:xt_proc} and  \ref{fig:lcurve}. 
According to the distance--time ($Y-T$) maps in Fig.~\ref{fig:xt_proc}
and time series curves in Fig.~\ref{fig:lcurve}, the enhancement in intensity is associated with
a Doppler blue-shift. Thus, these disturbances are consistent with an upward propagating
 slow magneto-acoustic wave \citep{2009ApJ...696.1448W}. Moreover, a projected propagation speed of about 
$60\pm 4.8$~km~s$^{-1}$ is sub-sonic (the sound speed C$_{S}\approx120$~km~s$^{-1}$ at the Ne~{\sc viii} 770~\AA\
 formation temperature of 630~000~K).

If these propagating disturbances were due to periodic plasma upflows then
we would expect that the highly blue-shifted velocity (about 10 to 18~km~s$^{-1}$ along the LOS for a radial
 structure making an angle of about $10\degree$ with the plane of sky at these latitudes for upflows between
 60 to 100~km~s$^{-1}$) associated to such upflow be on top of the average,
slightly blue-shifted, continuous outflow  typically seen in Ne~{\sc viii} in coronal holes.
At the lower latitudes at which the disturbances are seen, the continuous outflow speeds are small
and true red-shifts are observed at some phases of the wave.
This finding supports the notion that these propagating disturbances correspond to waves.
 However, we warn that this result depends sensitively on
 velocity calibration, and the difficulty of EUV wavelength calibration reduces its reliability.

We also looked for signatures of these disturbances in the transition region (N~{\sc iv} 765~\AA) 
to find indications of the physical event and source responsible for these disturbances. 
We could not find any propagating signature in any of the line parameters of N~{\sc iv} 765~\AA. 
However, the line width of N~{\sc iv} 765~\AA\ is enhanced at the location from where these 
propagating disturbances in Ne~{\sc viii} 770~\AA\ originate (bottom panel of Fig.~\ref{fig:xt_proc}).
More precisely, patches of enhanced width of the N~{\sc iv} 765~\AA\ line appear to correspond 
to the blue-shifted phase of the Ne~{\sc viii} 770~\AA\ line (see Fig.~\ref{fig:xt_proc}).
This enhanced width at lower height indicates either unresolved small-scale reconnections associated with
 explosive events \citep{1997Natur.386..811I} happening at that place as evident from associated
line width enhancement and Doppler blueshift of N~{\sc iv} 765~\AA\ profiles, or a local increase in turbulence.
 Thus, it may be possible that these propagating
 disturbances are triggered by such small-scale events causing the increased line width of transition region lines.
However, also the reverse can be true, with a periodic burst of explosive events triggered by the waves,
as has been proposed by \citet{2004A&A...419.1141N}. Note that relatively small increase in line width suggests
that either these are rather weak explosive events, or that the enhancement in line width has a different cause. 
 
\begin{figure*}[htb]
 \centering
\includegraphics[width=15.cm]{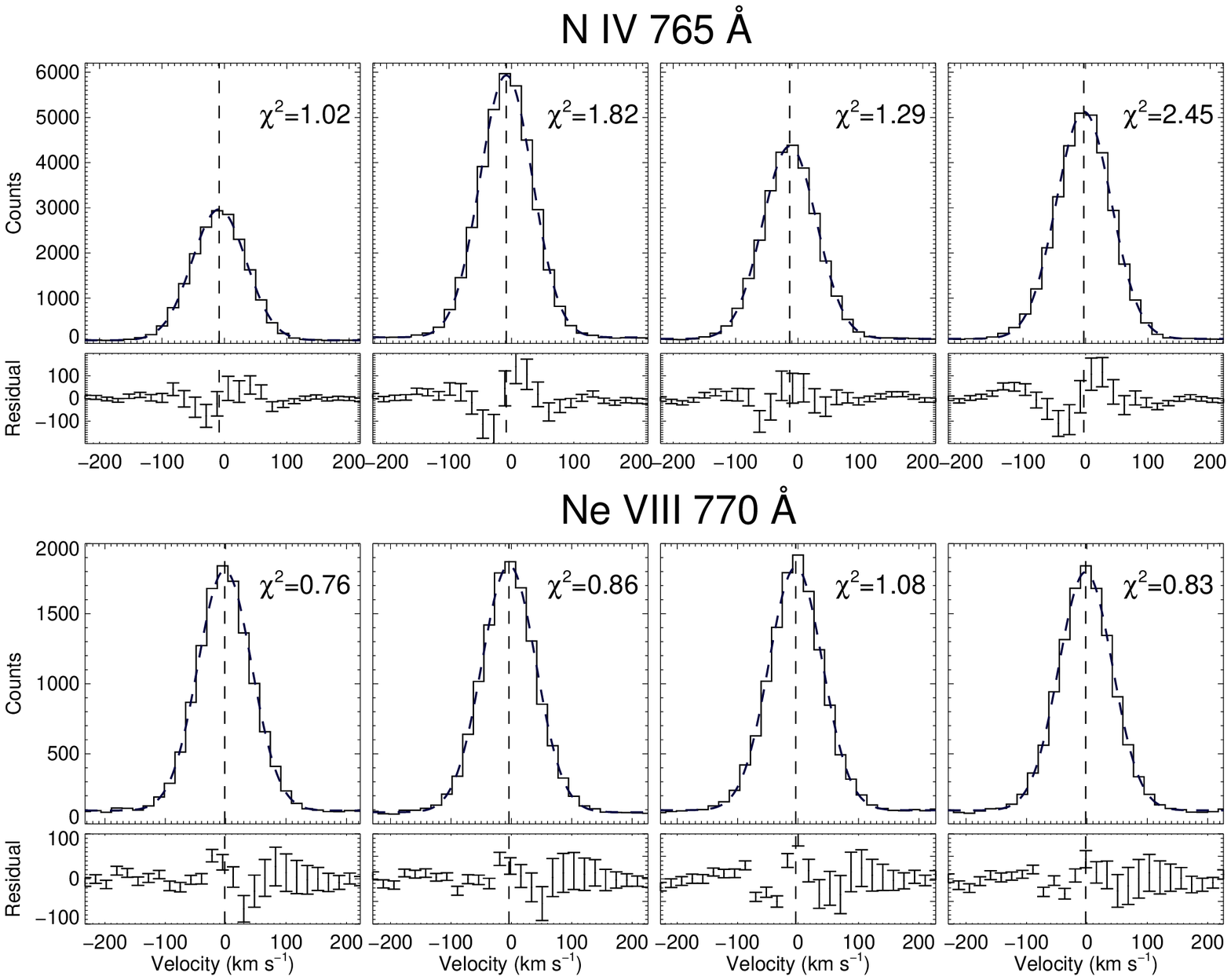}
\caption{Averaged spectral line profiles of N~{\sc iv} 765~\AA\ at the root of the propagating 
disturbances (top panels) at the four epochs of enhanced width in N~{\sc iv} (see bottom panel
 of Fig.~\ref{fig:xt_proc}) and (bottom panels) over the top quarter of the 
corresponding blue-shifted ridges in Ne~{\sc viii} (see second panel of Fig.~\ref{fig:xt_proc}). 
Profiles are plotted in the order of their occurrence in time. Both the N~{\sc iv} 765~\AA\ and 
Ne~{\sc viii} 770~\AA\ observed spectral profiles (solid histograms) are fitted with a single
 Gaussian component and constant background (dashed lines). The bars in the residual panels
 indicate the uncertainties of the measurement. Velocity scales are calibrated
 with respect to averaged off-limb wavelength. Vertical dashed lines in all panels mark the
position of the fitted line centroid.  }
\label{fig:line_prof}
\end{figure*}

To check the influence of the propagating disturbances on the spectral line profiles, 
in the upper row of Fig.~\ref{fig:line_prof} we plot the averaged line profiles of N~{\sc iv} 765~\AA\
at the root of the propagating disturbances, i.e. at the locations at which  
N~{\sc iv} 765~\AA\ displays enhanced line width (the plotted profiles are summed over
the areas marked with white ellipses in the bottom panel of 
Fig.~\ref{fig:xt_proc}). In the lower row of Fig.~\ref{fig:line_prof}, we plot Ne~{\sc viii} 770~\AA\
profiles summed  along the top quarter (low-latitudes) of the propagating ridge, 
where enhancements in intensity and blue-shifted velocities are clearly visible in Ne~{\sc viii} 770~\AA. 
The resulting profiles are ordered according to increasing time  
from left to right in Fig.~\ref{fig:line_prof}. 
Each plotted profile is fitted with a single Gaussian and a constant background. 
The fit residuals are also plotted. 
The average reduced chi-square, $\chi_{red}^{2}$, of four fits at the root of the propagating 
disturbances (in N~{\sc iv} 765~\AA) is about 1.65 and, together with the relatively consistent
 shape of the residual in all four profiles indicate
the presence of a small blue-shifted component (confirmed by a double Gaussian fit),
 which is particularly evident in the fourth panel from the left.
The average $\chi_{red}^{2}$ of about 0.88 and shapes of the residuals together show, however, that the 
Ne~{\sc viii} profiles averaged over the blue-shifted ridges are well represented by a single
Gaussian and leave little room for a significant second component. 
The presence of only one component may indicate that in this 
observation the background emission is weak, and most of the emission
 is coming from the oscillating coronal structure.


In general, it is very difficult to uniquely fit two Gaussian components to all spectral profiles for 
the whole sequence. Such fits have multiple solutions, with the final solution depending on the initial guess 
of the line parameters and also on the constraints applied to the fit. Thus, multi-Gaussian fitting with a 
unique set of initial guess parameters and constraints do not lead to any conclusive result.

To further check for the presence of a second component, we also applied the Red-Blue ($R-B$)
 asymmetry analysis to all the Ne~{\sc viii} 770~\AA\ spectral 
line profiles. After obtaining the line centroids from a single Gaussian fit, the line profiles
have been interpolated using linear and spline methods to a spectral spacing 10 times finer than the original one. 
We calculated the $R-B$ asymmetry at different velocities ($V_{RB}$) measured from the respective line centroid 
by using the formula \citep[after ][]{2010ApJ...722.1013D},

\begin{equation}
R-B=\frac{I(V_{RB})-I(-V_{RB})}{I(V_{RB})+I(-V_{RB})}
 \label{eq:r-b}
\end{equation} 
Here we integrated the intensity ($I$) of the interpolated profile over a narrow spectral range 
($\approx\pm1.7$~km~s$^{-1}$) at velocity ($V_{RB}$). Values of the ratio in Eq.~\ref{eq:r-b} 
give asymmetries whose sign fluctuates at different $V_{RB}$ and from profile to profile, indicating no clear
preference for a stronger red and blue wing.  
The $R-B$ asymmetry maps did not reveal propagating signatures in any of the velocity bins.
Yet individual line profiles, such as along the ridges of the propagating disturbance, 
still show a $R-B$ asymmetry with maximum amplitude of $\pm 0.07$ for $V_{RB} \leq 150$~km~s$^{-1}$. 
\citet{2010ApJ...724L.194V} pointed out the blueward
 and redward asymmetry in the line profiles at crest and trough epochs of wave propagation, respectively.
However, in our dataset we do not find any such preferred asymmetry in the line profiles at any point 
or epoch. 

\begin{figure*}[htb]
 \centering
\includegraphics[width=12.cm]{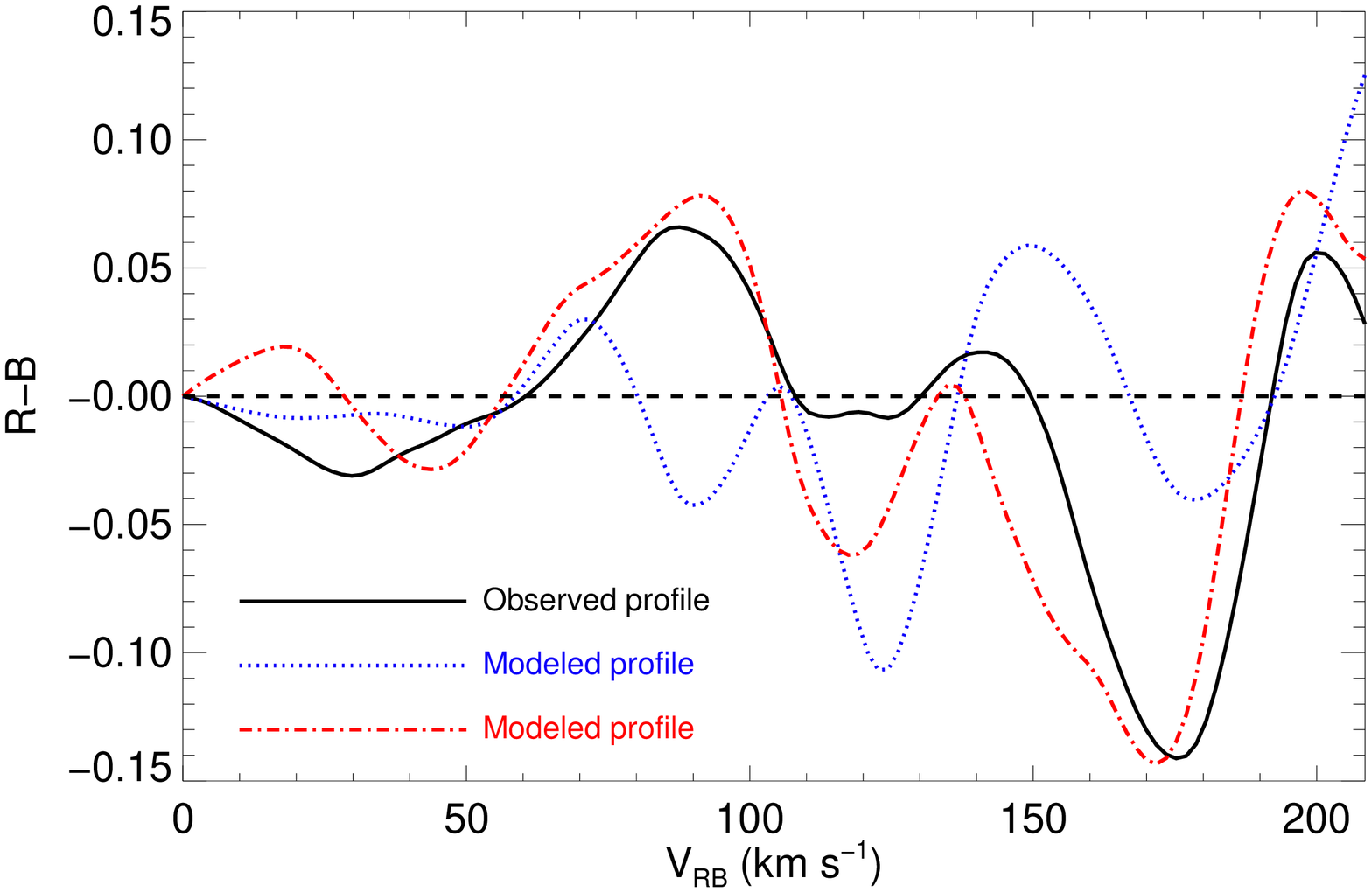}
\caption{Variation of Red-Blue ($R-B$) asymmetry of line profiles with velocities measured from
the respective line centroid ($V_{RB}$). The continuous black curve is obtained from the observed profile
 of Ne~{\sc viii} 770~\AA\ shown in the first bottom panel of Fig.~\ref{fig:line_prof}. 
The red and blue curves are obtained from 
the simulated profile modeled with a single Gaussian and a constant background (with parameters similar 
to the observed profile) but sampled differently with respect to the line centroid.
The red curve represents the case of maximum asymmetry in sampling (about half a pixel shift from the line centroid)
 while the blue curve is sampled symmetrically and, as such, shows just the effects of noise.}
\label{fig:comp_rb}
\end{figure*}

To test whether the resultant $R-B$ asymmetry values are significant and of solar origin,
we performed a similar analysis on simulated line profiles.
We generated a symmetric line profile (single Gaussian with constant background) with 
amplitude and FWHM similar to those in our dataset, but with 
spectral resolution ten times finer than the SUMER resolution. 
 Finally, we added random noise to the signal in each data
point, assuming a Poisson distribution. From this high-resolution profile, we extracted 
several profiles with spectral sampling similar to SUMER by sampling every tenth data point 
with different starting points. 
Finally, we performed an identical $R-B$ analysis on these simulated line profiles as on
the observations. The resultant $R-B$ asymmetry profile shows Red-Blue wing asymmetries 
 with amplitudes similar to those obtained from the observed profiles (see Fig.~\ref{fig:comp_rb}),
 although the modeled profile was completely symmetric, only sampled unsymmetrically .  
 The effect apparently arises from the finite sampling of the
data and is very susceptible to noise.
In fact, the amplitude of the asymmetry increases with velocity as the fractional error in 
the data points increases when moving away from the line centroid. 
The $R-B$ asymmetry profile obtained from this analysis is similar to the $R-B$ asymmetry profile 
obtained from the real data in terms of amplitude and sign of asymmetry 
(maximum amplitude is about $\pm 0.08$ for $V_{RB} \leq 150$~km~s$^{-1}$ for the modeled profiles,
see Fig.~\ref{fig:comp_rb}).
 Thus, we conclude that within the given signal-to-noise ratio and spectral sampling of our data,
 $R-B$ asymmetry analysis can not detect any reliable secondary component.
This finding supports the result that the Ne~{\sc viii} 770~\AA\ line profiles
 associated with propagating disturbances are essentially symmetric in nature.  


\citet{2010ApJ...722.1013D} have shown that a single Gaussian fit to a spectral 
profile, having a quasi--periodically varying additional faint component of blue-shift emission, 
can result in quasi--periodic
modulation of the peak intensity, centroid and width of a line, and thus could produce a small oscillatory 
signature in these line parameters. However, in our analysis we find the Doppler velocity amplitude 
of these disturbances to be relatively large at low latitudes, i.e., $\pm$3~km~s$^{-1}$ (on top of the average
plasma outflow speed), which is much larger than that resulting from the simulations
(about 1~km~s$^{-1}$) of \citet{2010ApJ...722.1013D}. In fact, the above simulations could produce larger
 Doppler shifts by assuming a relatively large second component that would, however, contradict the rather
 small intensity increases we observe.
\citet{2011ApJ...727L..37T}  looked at the base
of a loop  and found in-phase intensity, Doppler velocity and line-width 
variations and interpreted those to be due to fast flows.
\citet{2011ApJ...737L..43N} also looked at greater heights from the same loop foot-point
and interpreted the propagating disturbances in terms of slow magneto-acoustic waves
based on similar large-amplitude Doppler velocity oscillations (5~km~s$^{-1}$ peak-to-peak) 
with a noticeably correlated intensity and Doppler velocity and lack of evidence of blueward asymmetry
 in the line profiles.
In our observations, the results are very similar to those obtained by 
\citet{2011ApJ...737L..43N}, and thus appear to support the conventional interpretation 
of these disturbances in terms of slow magneto-acoustic waves, as was done by 
\citet{2011ApJ...734...81M} based on the measured sub-sonic speed of propagation.
Moreover, a recent 3D MHD model of hot coronal loops developed by \citet{2012ApJ...754..111O} shows the
 close relationship between slow waves and impulsive upflows at the loop foot-points which were generated
 by the same impulsive events. Thus, in order to distinguish between slow waves and quasi-periodic upflows,
 a detailed analysis of the relationship between different oscillation parameters (e.g. period, phase speed, 
phase relations, damping time and length) and loop parameters (e.g. loop length, density, temperature
 and magnetic field) are required as demonstrated and pointed out by \citet{2012ApJ...754..111O}.

\section{Conclusion}
 
In summary, we found presence of propagating disturbances in a polar coronal hole as 
revealed by intensity, velocity and line width of the Ne~{\sc viii} 770~\AA\ spectral line,
 with a projected propagation speed of $60 \pm 4.8$~km~s$^{-1}$ and a period $\approx$14.5~min.
The observed disturbances show quasi-periodic enhancement in intensity in phase with
the Doppler blue-shift. At lower latitudes, true red-shifts are also observed at some
phases of the propagation.
We studied Ne~{\sc viii} spectral line profiles averaged along the propagating ridges
and found them to be symmetric, to be well fitted by a single Gaussian, and to have no noticeable 
Red-Blue asymmetry. Based on our observations and analysis, we conclude that the most likely
 cause for these propagating disturbances in coronal hole regions are slow magneto-acoustic waves.
Observations at lower heights in the solar atmosphere with the N~{\sc iv} 765~\AA\ spectral line
do not show any propagating features in any line parameter. However, provides the indication that
these waves may be caused by small-scale reconnections or explosive events occurring at the base 
of the propagating features.
 With the given propagation speed and velocity amplitude of these waves, the energy flux carried by
them is far less than that needed for the heating of the solar corona 
\citep{2009ApJ...696.1448W,2010ApJ...721..744K}. However, these waves may provide additional
 momentum for the acceleration of the fast solar wind in coronal holes \citep{2010LRSP....7....4O}
and their dissipation could also lead to extended heating of the outer coronal holes.

\begin{acknowledgements} 
This work was partially supported by the Indo-German DST-DAAD
 joint project D/07/03045. 
The SUMER project is financially supported by DLR, CNES, NASA, and the ESA  
PRODEX programme (Swiss contribution). 
This work was partially supported by the WCU grant No. R31-10016 from the 
Korean Ministry of Education, Science and Technology.
\end{acknowledgements}

\bibliographystyle{aa.bst}
\bibliography{/home/girjesh/research/papers/references}

\end{document}